\documentclass[prc,twocolumn,floatfix,nofootinbib]{revtex4-1}

\usepackage{graphicx}
\usepackage{amssymb,amsmath,amstext,amsthm,amsfonts}
\usepackage{slashed}

\begin{document}

\title{Reaction Mechanisms at MINER$\nu$A}

\author{U. Mosel}
\email[Contact e-mail: ]{mosel@physik.uni-giessen.de}
\author{O. Lalakulich}
\affiliation{Institut f\"ur Theoretische Physik, Universit\"at Giessen, D-35392 Giessen, Germany}
\author{K. Gallmeister}
\affiliation{Institut f\"ur Theoretische Physik, Johann Wolfgang Goethe-Universit\"at, D-60438 Frankfurt a.\ M., Germany}

\begin{abstract}

The MINER$\nu$A experiment investigates neutrino interactions with nucleons needed for an understanding of electroweak interactions of hadrons. Since nuclear targets are being used many-body effects may affect the extracted cross sections and the energy reconstruction. The latter is essential for the extraction of neutrino oscillation properties. We investigate the influence of nuclear effects on neutrino interaction cross sections and make predictions for charged current quasielastic (QE) scattering, nucleon-knock-out and pion- and kaon-production on a CH target. The Giessen Boltzmann--Uehling--Uhlenbeck (GiBUU) model is used for the description of neutrino-nucleus reactions. Integrated and differential cross sections for inclusive neutrino scattering, QE processes and particle production for the MINER$\nu$A neutrino flux are calculated. The influence of final state interactions on the identification of these processes is discussed. In particular, energy and $Q^2$ reconstruction for the MINER$\nu$A flux are critically examined. The $Q^2$ dependence of the inclusive cross sections is found to be sensitive to the energy reconstruction. Cut-offs in flux distributions have a large effect. Final state interactions affect the pion kinetic energy spectra significantly and increase the kaon cross sections by cross feeding from other channels.

\end{abstract}

\date{\today}

\maketitle

\section{Introduction}
Neutrino cross sections and nuclear effects can lead to systematic uncertainties in the extraction of neutrino oscillation parameters. The MINER$\nu$A experiment, therefore,  aims to investigate both of these in a neutrino beam whose flux is similar to that of the planned Long Baseline Neutrino Experiment (LBNE). Precision measurements of the differential cross sections for quasielastic scattering (QE) and single- and multi-pion production, as a function of (reconstructed) neutrino energy, are expected to reduce uncertainties in the tuning of generators \cite{Minerva:2005}. These are ultimately needed to extract the oscillation parameters from the measured event distributions. In addition, also studies of neutrino-induced $\Delta S = 0$ strangeness production are planned \cite{Solomey:2005rs}. First results from MINER$\nu$A have recently been published \cite{Fiorentini:2013ezn,Fields:2013zhk}.

Over the last few years valuable insight into the neutrino-nucleus interactions has been obtained mainly from the MiniBooNE experiment which has reported cross sections for QE-like events, for reconstructed QE events \cite{AguilarArevalo:2010cx}, for inclusive cross sections \cite{AguilarArevalo:2010zc}, double-differential in the outgoing muon variables, and for pion production \cite{AguilarArevalo:2010bm,AguilarArevalo:2010xt}. It came as a surprise that the cross section originally identified as quasielastic contains a $\approx 30\%$ contribution from so-called 2p-2h events in which the incoming neutrino couples to two nucleons \cite{Martini:2009uj,Martini:2010ex,Martini:2010ex,Martini:2011wp,Nieves:2011pp,Nieves:2011yp}.

The original misidentification has led to incorrect cross section determinations and to the extraction of incorrect physical parameters (in this case the axial mass) by tuning the generators. It also has far reaching consequences for the energy reconstruction. This has been discussed in recent Refs. \cite{Martini:2012fa,Martini:2012uc,Nieves:2012yz,Lalakulich:2012hs,Benhar:2013bwa} for the MiniBooNE experiment. In \cite{Lalakulich:2012hs,Martini:2012uc} the influence of the errors inherent in the reconstruction procedure on the oscillation signal observed in T2K was discussed (see also \cite{Meloni:2012fq,Coloma:2013rqa,Coloma:2013tba}).

For the MINER$\nu$A flux with its higher energies equally important is the process of pion production through resonance excitations or deep inelastic scattering (DIS). Here the MiniBooNE data have created a new puzzle: the calculated integrated cross sections for single pion production are overall, but mainly at the higher neutrino energies, somewhat smaller than the measured ones. This is so even when the larger elementary data from an old BNL experiment are used (for a discussion see \cite{Lalakulich:2012cj}). Furthermore, the calculated pion momentum-spectra exhibit a distinct dip for momenta corresponding to the excitation of the $\Delta$ resonance, as a consequence of pion reabsorption through this resonance \cite{Lalakulich:2012cj,Hernandez:2013jka}; this dip is not present in the published data \cite{AguilarArevalo:2010bm,AguilarArevalo:2010xt}. Any new data on pion production from the MINER$\nu$A experiment
could thus help to clarify the reasons for these problems.

In \cite{Lalakulich:2012gm} we have published the results of extensive  calculations of QE scattering and particle production cross sections for the MINOS and NO$\nu$A experiments for Iron (Fe) and Carbon (C) targets, respectively. As an addendum to that work we present here now results for the MINER$\nu$A experiment, using its flux and a CH target. The main aim of this study is to shine some light on energy- and $Q^2$ reconstruction in the higher energy regime of the MINER$\nu$A experiment and to make predictions for nucleon-knockout, pion- and kaon-production.

\section{Method}

We use the transport model GiBUU to model both the initial and the final state interactions \cite{Buss:2011mx,gibuu}. This model has been widely used and tested in very different physics scenarios, ranging from a description of heavy-ion reactions to photon-, electron- and neutrino-induced reactions. It combines the initial reaction mechanisms true QE scattering, 2p-2h excitations and pion (and other meson) production through resonances, background processes and deep inelastic scattering (DIS) with an extensively tested description of final state interactions \cite{Buss:2011mx}.

The nucleons move in a location- and momentum-dependent potential well; their spectral function is that of a bound, local Fermi-gas. The description of QE scattering uses an axial mass $M_A = 1$ GeV. Pion production is in its vector interaction part consistent with the MAID analysis for electron scattering \cite{MAID} for nucleon resonances up to an invariant mass of about 2 GeV.  The axial part is constrained by effective field theory at least in its low-energytransfer region \cite{Lalakulich:2010ss} and by PCAC, assuming dipole form factors, for the higher lying nucleon resonances; the coupling strength to the $\Delta$ is obtained from fitting to the BNL data \cite{Kitagaki:1986ct}.  DIS is treated via the high-energy hadronization model \textsc{PYTHIA} \cite{Sjostrand:2006za}. No parameters in GiBUU are tuned to nuclear neutrino data, with one exception: the strength of the genuine two-body contribution of the elementary 2p-2h hadron tensor. We note here first that final state 2p-2h contributions of the type $N + N \to N + N$ or $\Delta + N \to N + N$ are explicitly included in the GiBUU transport. The extra two-body current contribution to the 2p-2h excitation for $N + N \to N + N$  is modeled to be purely transverse, with a $Q^2$ dependent dipole form factor with a cut-off parameter of $\Lambda = 1.5$ GeV; the nuclear contribution of this term is in addition affected by nuclear binding, Fermi-motion, Pauli-blocking and final state interactions (fsi) of the two nucleons\footnote{This 2p-2h contribution was not present in the results presented earlier for the MINOS and NO$\nu$A experiments \cite{Lalakulich:2012gm}.}. Its strength is fitted to the semi-inclusive MiniBooNE double-differential no-pion data \cite{Lalakulich:2012ac} and as such contains both the effects of 2p-2h excitations and of RPA correlations known to be essential at small $Q^2$.  For all further details we refer to \cite{Buss:2011mx,gibuu,Lalakulich:2012ac,Lalakulich:2012gm}.

All calculations are done for a CH target using the MINER$\nu$A flux for the energy window from 1.5 to 10 GeV  \cite{Dytmanprivcom:2013}; the latter corresponds to that used in the experimental analysis.

\section{Results}
\subsection{Energy and $Q^2$ Reconstruction}
In \cite{Fiorentini:2013ezn} the MINER$\nu$A collaboration has published an analysis of QE scattering cross sections as a function of $Q^2$ and has compared that to predictions of event generators using various recipes for the approximate treatment of 2p-2h interactions. Since $Q^2$ cannot directly be measured, it has been reconstructed from the kinematical variables of the outgoing muon using the following two equations
\begin{eqnarray}    \label{E_Q2}
E^{\rm rec}_{\nu} &=& \frac{2(M_n - E_B)E_\mu - (E_B^2 - 2 M_nE_B + m_\mu^2 + \Delta M^2)}{2(M_n - E_B - E_\mu + |\vec{k}_\mu| \cos \theta_{\mu})}
\nonumber \\
Q^2_{\rm rec} &=& 2 E_{\nu}^{\rm rec} (E_\mu - |\vec{k}_\mu| \cos \theta_{\mu}) - m_\mu^2 ~,
\end{eqnarray}
which are correct for QE scattering on a neutron at rest. Here $\vec{k}_\mu$ is the outgoing lepton's momentum, $E_\mu$ its energy and $\theta_\mu$ its angle. The quantity $E_B$ denotes an average binding energy of the neutron inside the nucleus; it is taken to be $E_B = 0.03$ GeV. Furthermore, $\Delta M^2 = M_n^2 - M_p^2$.
For our analysis we assume GiBUU to be 'nature' and generate full events for true energies distributed according to the MINER$\nu$A flux. Although the true energy and momentum transfer are known for each event we also reconstruct both of these quantities using Eq.\ (\ref{E_Q2}).

The experimental analysis identifies QE scattering by requiring an outgoing muon in the final state along with one or more nucleons and no mesons \cite{Fiorentini:2013ezn}; in the following discussion these will be called 0-pion events.
It is well known that such events do contain so-called 'stuck-pion' events in which a pion is first produced but subsequently absorbed. In \cite{Leitner:2010kp} we have shown that such events can lead to errors in the reconstructed energies in addition to the smearing to be expected from Fermi motion of the neutrons in the nuclear target. In \cite{Lalakulich:2012hs,Martini:2012fa,Martini:2012uc,Nieves:2012yz,Benhar:2013bwa} it has been shown that also 2p-2h excitations have a similar effect. All these complications are, therefore, also expected to be present in the results to be shown here. In addition, pion production and DIS events are expected to play a much larger role at the energies of the MINER$\nu$A experiment than at the significantly lower energies of MiniBooNE and T2K.

\begin{figure}[htb]
 \includegraphics[angle=-90,width=0.5\textwidth]{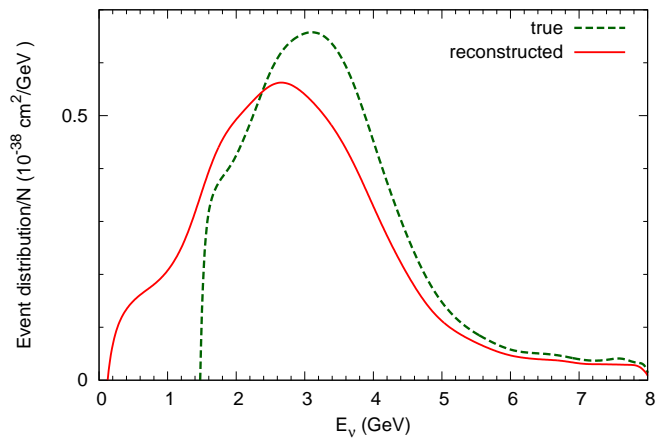}
 \caption{(color online) 0-pion event distribution (flux times cross section) per neutron for MINER$\nu$A vs.\ true (dashed green curve) and reconstructed (solid red curve) energy.} \label{fig:ev_no_pion}
\end{figure}

We start our discussion by showing first  in Fig.\ \ref{fig:ev_no_pion} the event distribution $\Phi(E_\nu) \sigma(E_\nu)$ for the 0-pion events, both as a function of true and of reconstructed energy (the latter has been obtained from Eq.\ (\ref{E_Q2}) as in the experimental analysis). $\Phi(E_\nu)$ is the energy distribution of incoming neutrinos, normalized to 1, and $\sigma(E_\nu)$ is the total cross section at energy $E_\nu$ for all events with a muon, 0 pions, and any number of nucleons, in the outgoing channel.
It is seen that there is a large difference between the true and the reconstructed event distributions. Particularly impressive is the significant strength that the latter shows for neutrino energies below 1.5 GeV, where no incoming true flux exists. This extra strength is removed from the peak region of the flux distribution where the effects of the low true energy cutoff at 1.5 GeV propagate up to about 6 GeV. The shift at the peak amounts to about 400 MeV. It is thus larger than that observed at the lower energies of the MiniBooNE and T2K experiments \cite{Lalakulich:2012hs} reflecting the presence of more resonance excitations and DIS at these higher energies. At still higher energies, above about 6 GeV, the two distributions coincide thus making the extraction of energy-dependent cross sections less dependent on any high-energy cut-off (at 10 GeV in the experiment and the present calculation). That the reconstruction is so much more sensitive to the low-energy cutoff than to the one at high energy is due to the flux distribution: while the flux is already at about 50\% of its maximum at $E_\nu = 1.5$ GeV it has fallen to about 3\% at the high-energy cutoff of 10 GeV.

\begin{figure}[htb]
 \includegraphics[angle=-90,width=0.5\textwidth]{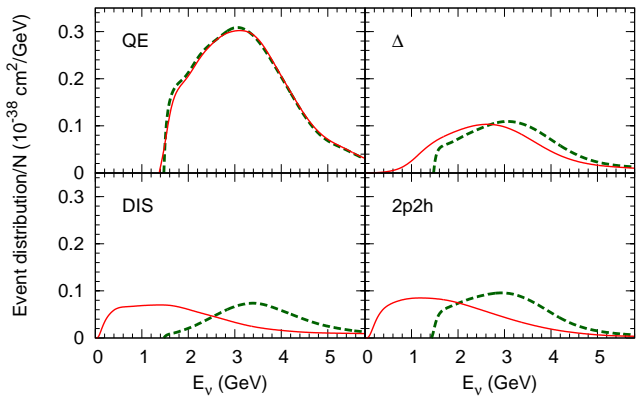}
 \caption{(color online) Individual 0-pion event distributions (flux times cross section) per neutron in the MINER$\nu$A flux vs.\ true (dashed green curve) and reconstructed (solid red curve) energy. The primary interaction processes are indicated in the figure: QE denotes the contribution from quasielastic scattering, $\Delta$ that from excitation of the $\Delta$ resonance, DIS that from deep inelastic scattering and 2p2h that from 2p-2h interactions.}
 \label{fig:ev_no_pion_multi}
\end{figure}
At the lower energies of the MiniBooNE and T2K experiments the downward shift of reconstructed energies was mainly due the presence of 2p-2h excitations \cite{Lalakulich:2012hs,Martini:2012fa,Martini:2012uc,Nieves:2012yz,Benhar:2013bwa}. Fig.\ \ref{fig:ev_no_pion_multi} shows the breakdown of the observed shift in the event rates by the most important primary reaction channels. It is seen that the reconstruction for the true QE process (top left in Fig.\ \ref{fig:ev_no_pion_multi}) works very well; the small discrepancies seen are due to the presence of a location- and momentum-dependent potential for the outgoing nucleon. The other three reaction mechanisms all exhibit a downward shift in the event rates vs.\ reconstructed energy. The shifts for 2p-2h, $\Delta$ excitation and DIS are comparable in magnitude; this is new compared to the case at the lower energies where DIS played no significant role. Further, smaller contributions (not shown here) come from pion background production and excitation of higher nucleon resonances.

\subsection{$Q^2$ Distributions}
In \cite{Fiorentini:2013ezn} the extracted $Q^2$ distribution was compared with various generator options (increased axial mass and transverse enhancement in NuWro \cite{Nuwro} compared to the standard result obtained in GENIE \cite{genie}). The calculations used for this comparison were made for a pure QE event sample thus neglecting any complications in the reconstruction that might arise from a misidentification of the QE reaction mechanism. It is therefore interesting to also look at this distribution here. According to Eq.\ (\ref{E_Q2}) the $Q^2$ distribution depends on $E_{\nu}^{\rm rec}$; thus any errors made in the determination of the energy directly propagate into an error of $Q^2$.

\begin{figure}[h]
\includegraphics[angle=-90,width=0.5\textwidth]{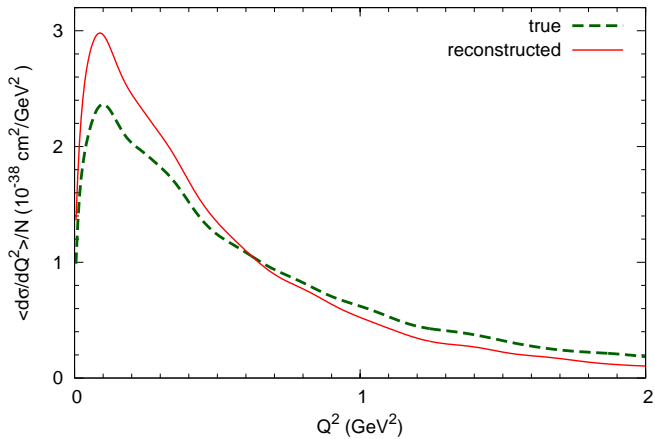}
\caption{(color online) Flux-averaged cross section $d\sigma/dQ^2$ per neutron as a function of true (dashed, green) and reconstructed (solid, red) $Q^2$. For both curves only 0-pion events are considered.} \label{fig:Q2rec}
\end{figure}

We, therefore, now turn to a discussion of the flux averaged $Q^2$ distribution
\begin{equation}
\langle d\sigma/dQ^2 \rangle = \int \Phi(E_\nu) \frac{d\sigma}{dQ^2}(E_\nu) \,dE_\nu   ~.
\end{equation}
Here $\frac{d\sigma}{dQ^2}(E_\nu)$ is the $Q^2$-differential cross section at neutrino energy $E_\nu$.
The corresponding flux-averaged $\langle d\sigma/dQ^2 \rangle$ is shown in Fig.\ \ref{fig:Q2rec} both as a function of true and of reconstructed $Q^2$.
Up to $Q^2 \approx 0.6$ GeV$^2$ the reconstructed $Q^2$ distribution is higher than the true one; at the peak at $Q^2 \approx 0.1$ GeV$^2$ it is about 25\% higher than the true distribution. Also the slope is higher for $Q^2$ values between about 0.1 and 1.5 GeV$^2$; this is just the $Q^2$ range which played an important role in the analysis of \cite{Fiorentini:2013ezn}. In terms of an axial mass extracted from the slope of $d\sigma/dQ^2$ the curve vs.\ reconstructed $Q^2$ corresponds to a smaller value of $M_A$ than the true one.

\begin{figure}
\includegraphics[angle=-90,width=0.5\textwidth]{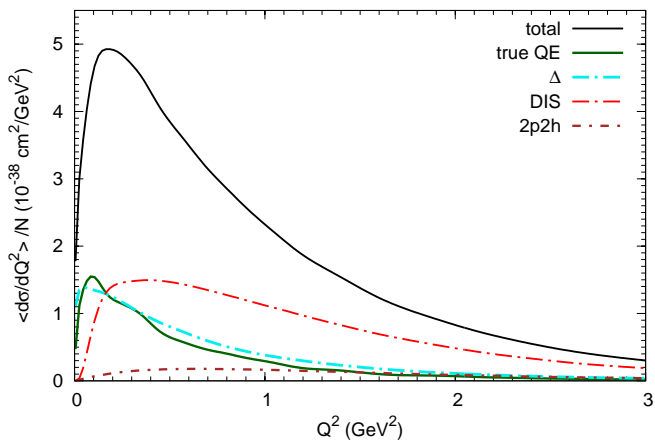}
\caption{(color online) Flux-averaged cross section $d\sigma/dQ^2$ per neutron for all events as a function of true $Q^2$. The individual curves give the cross section for some primary reactions, as indicated in the figure.} \label{fig:Q2}
\end{figure}

In Fig.\ \ref{fig:Q2} we show a breakdown of the true $Q^2$ distribution into some of the most important primary, initial interaction process for all events.
It is seen that the total cross section (upper black solid line) is about 3.5 times larger than the true QE cross section (lower solid dark-green line). The $\Delta$ contribution is as large as that of true QE and similar in shape. However, the effect of the Pauli-principle, which shows up in the suppression at very small $Q^2$, is less pronounced for the $\Delta$ than for the true QE contribution. This is due to the fact that a large part of the $\Delta$ contribution is due to pionless decay in the nuclear medium $\Delta + N \to N + N$. This process provides enough energy to kick the final state nucleon above the Fermi surface. The DIS contribution becomes dominant for  $Q^2 > 0.2$ GeV$^2$ and is overall the largest of these three components. The 2p-2h component is of minor importance compared to all these other components. The figure shows that indeed at MINER$\nu$A true QE scattering is not a dominant contribution (about 1/3) of the total cross section.

As in the experimental analysis we now impose the condition of 0 pions in the final state.
\begin{figure}[htb]
\includegraphics[angle=-90,width=0.5\textwidth]{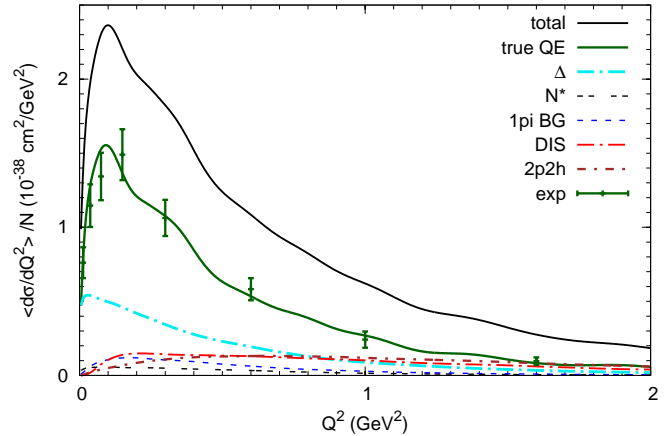}
\caption{(color online) Flux-averaged cross section $d\sigma/dQ^2$ per neutron for 0-pion events as a function of true $Q^2$. The individual curves give the cross section for various primary reactions, as indicated in the figure. The primary reaction labels are as in Fig.\ \ref{fig:ev_no_pion_multi}. In addition, $N^*$ denotes contributions from higher-lying
nucleon resonances and 1pi BG denotes those from pion production background events (Born diagrams). The data are taken from \cite{Fiorentini:2013ezn}.} \label{fig:Q2nopi}
\end{figure}
The results are shown in Fig.\ \ref{fig:Q2nopi}; for completeness we show here also the smaller production channels not contained in Fig. \ref{fig:Q2}. The main effect of this condition is to remove a large part of the DIS and $\Delta$ cross sections. Now the total is at its peak only about 1.5 times larger than the true QE contribution; QE events have effectively been enriched. After the restriction to 0-pion events, at small $Q^2$ true QE and $\Delta$ contributions dominate. All other contributions, and in particular the strong DIS contribution, are strongly suppressed there. That the primary $\Delta$ contribution still remains reasonably strong below about $Q^2 = 0.6$ GeV$^2$ is due to the  pionless decay in medium.

In Fig.\ \ref{fig:Q2nopi} we also show the data points taken from \cite{Fiorentini:2013ezn}. These data were obtained from the 'raw' data by subtracting various, substantial background contributions to the cross section, as obtained from the GENIE generator; they thus contain some generator dependence. Overall, the data are described quite well by the present GiBUU calculation for the true QE component (solid green curve in Fig.\ \ref{fig:Q2nopi}). Possible discrepancies show up around the peak of the $Q^2$ distribution, i.e.\ at small $Q^2 \approx 0.075 {\rm GeV}^2$. This is the region where the errors in the $Q^2$ reconstruction are largest (see Fig.\ \ref{fig:Q2rec}). Also here a sizeable contribution from $\Delta$ excitation with subsequent pion reabsorption exists in the 0 pion event sample (see blue dashed-dotted curve in Fig.\ \ref{fig:Q2nopi}). It is exactly this $\Delta$ contribution which is the most uncertain one among the various elementary processes \cite{Lalakulich:2012cj} so that the small discrepancy in the $d\sigma/dQ^2$ distribution could be easily removed by a small change of the overall $\Delta$ strength and/or a slight change in its form factor.

\subsection{Particle Production}

In the following three subsections we now turn to a discussion of particle production cross sections.

\subsubsection{Nucleon Knockout}
\begin{figure}[h]
\includegraphics[angle=-90,width=0.5\textwidth]{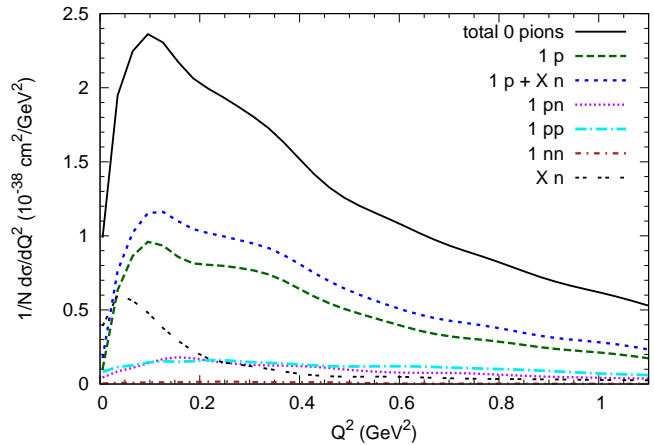}
\caption{(color online) Flux-averaged cross section $d\sigma/dQ^2$ per neutron for 0-pion events as a function of true $Q^2$. The individual curves give the cross section for various reaction channels, as indicated in the figure.} \label{fig:Q2observ}
\end{figure}

Whereas the discussions in the preceding section dealt with the breakdown of the observed $Q^2$ distribution into its various primary interaction contributions we now turn to a discussion of the same distribution in terms of observables. In Fig.\ \ref{fig:Q2observ} we show the contributions of various reaction channels, all with 0 pions, to the observed $Q^2$ distribution.
Both the processes with exactly 1 proton in the final state, and no other hadrons, and the one with 1 proton, X neutrons and no other hadrons, make up about 40\% of the total cross section, with very similar shapes. The channels with a pp or a pn pair are very similar, quite flat and suppressed and thus of minor importance. Interesting, however, is the pile-up of strength seen in the Xn channel at small $Q^2 \approx 0.1$ GeV$^2$. This is entirely due to fsi. The outgoing channel with 1 muon, 0 pions, exactly 1 proton and any number of neutrons is made up to 90\% by a true primary QE event, as we have verified by an explicit analysis. Selective observation of this channel would thus make the energy- and $Q^2$ reconstruction quite reliable, at the expense of losing about 1/3 of the QE cross section. This also holds for the LBNE \cite{Mosel:2013fxa}.

\begin{figure}
\includegraphics[angle=-90,width=0.5\textwidth]{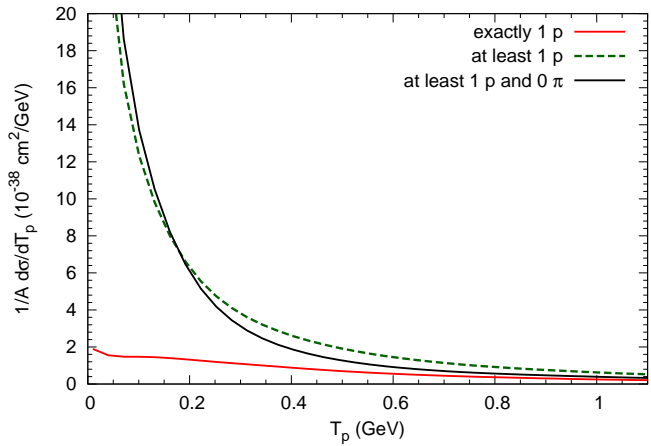}
\caption{(color online) Flux-averaged kinetic energy distribution per nucleon of knock-out protons. The upper dashed (green) curve gives the semi-inclusive cross section for events with at least 1 proton and any number of other hadrons with different charge, the lower solid (red) curve that for exclusive events with exactly 1 proton and no other hadrons. The upper solid (black) curve gives the semi-inclusive cross section for 0 pion events.} \label{fig:Nspectr}
\end{figure}

The kinetic energy spectrum for proton-knockout is shown in Fig.\ \ref{fig:Nspectr}. It is seen that for MINER$\nu$A the probability for a single proton knockout is quite small; the spectrum for that process is fairly flat.  Its cross section is by far overshadowed by that for an inclusive process, in which at least one proton and any number of other hadrons is observed. The spectrum for this semi-inclusive proton-knockout is peaked at small energies and falls of steeply with kinetic energy of the proton. This is a consequence of fsi in which the initial particle struck by the neutrino propagates through the nucleus and ejects more and more particles; energy conservation then leads to the predominance of low-energy protons. A closer analysis shows that the most dominant source of these semi-inclusive protons is DIS, which amounts to about 1/2 of the total at 200 MeV, to be followed by $\Delta$ excitation amounting to about 1/3 of the total. In both cases the relevant fsi reaction is that of pionless decay which knocks out a nucleon. All the other processes such as 2p2h, higher lying nucleon resonances and 1 pion background events account for the rest. While these results were obtained for all events, Fig.\ \ref{fig:Nspectr} also shows the semi-inclusive cross section for knock-out protons for 0 pion events only (upper black solid curve). One sees that now even more energy is redistributed from high energies to lower ones, with the crossing point at about 200 MeV. The energy carried by the pions in the full event is now to be found in the many nucleons with energies below about 200 MeV.
The spectrum of these final state protons falls continuously with energy so that it is not possible to give a threshold energy below which secondary nucleons dominate. It is clearly seen, however, that most of the cross section for 1p knockout resides at small kinetic energies

\subsubsection{Pion Production}
\begin{figure}
\includegraphics[angle=-90,width=0.5\textwidth]{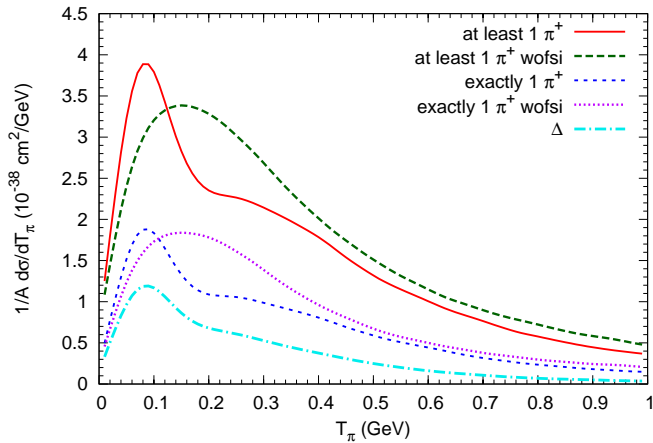}
\caption{(color online)  Flux-averaged kinetic energy distribution of $\pi^+$. The uppermost solid (red) curve gives the cross section per nucleon for events with at least 1 $\pi^+$ and any number of other hadrons, the lower dotted (blue) curve that for events with exactly 1 $\pi^+$ and no other hadrons. The other 2 curves give the corresponding results without any final state interactions. The lowest, dot-dashed magenta curve shows the contribution to single-pion events from the $\Delta$ resonance only.} \label{fig:pi+}
\end{figure}

\begin{figure}
\includegraphics[angle=-90,width=0.5\textwidth]{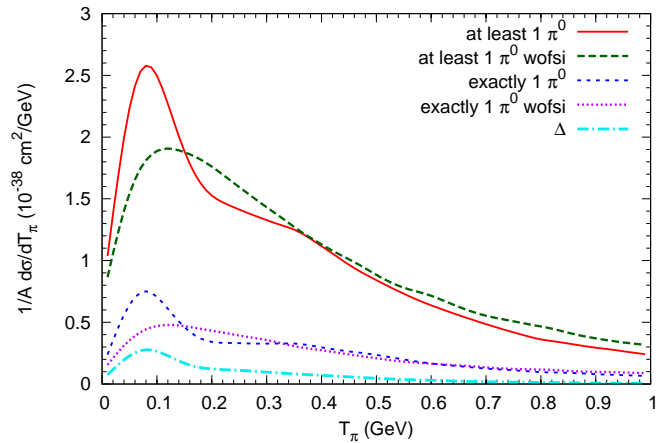}
\caption{(color online) Flux-averaged kinetic energy distribution of $\pi^0$. The uppermost solid (red) curve gives the cross section per nucleon for events with at least 1 $\pi^0$ and any number of other hadrons, the lower dotted (blue) curve that for events with exactly 1 $\pi^0$ and no other hadrons. The other 2 curves give the corresponding results without any final state interactions. The lowest, dot-dashed magenta curve shows the contribution from single-pion events from the $\Delta$ resonance only.} \label{fig:pi0}
\end{figure}

Pion production presents a major background to charged current QE processes. Pions can originate in three elementary reactions: through the $\Delta$ and higher resonances, through background processes (Born diagrams) and through DIS, with the first and the last being the dominant ones. In Figs.\ \ref{fig:pi+} and \ref{fig:pi0} we show the pion spectra for exclusive 1-pion production and semi-inclusive pion production. For both also the spectra at the end of the initial interaction are shown, without further final state interactions. The semi-inclusive cross sections are larger than the exclusive ones by factors of 2 - 5, for $\pi^+$ and $\pi^0$ production, respectively; this reflects the fact that the beam energy is high enough for multi-pion production.

For both charge states the figures show the dramatic impact of fsi. Whereas the spectra before fsi are broadly peaked around 0.15 GeV (for $\pi^+$), after fsi they look much more pointed with a peak at an energy of about 0.08 GeV. At the peak the cross sections after fsi are larger than those before; at the same time a dip in the spectrum develops at around 0.2 GeV. This behavior was already seen in the early calculations for pion spectra \cite{Leitner:2006sp,Leitner:2006ww}. It was later on confirmed in calculations for the MiniBooNE and T2K experiments \cite{Leitner:2008wx,Leitner:2009ec,Leitner:2009de,Lalakulich:2012cj,Lalakulich:2013iaa,Hernandez:2013jka}. This dip is a consequence of two competing effects: on one hand, pion absorption through the $\Delta$ resonance is most pronounced at $T_\pi \approx 0.13$ GeV; on the other hand, high energy pions get slowed down by elastic and inelastic interactions. Both effects are responsible for the build-up of the peak in the spectrum at 0.08 GeV that even exceeds the cross section  there before fsi. Figs.\ \ref{fig:pi+},\ref{fig:pi0} also show the pion's kinetic energy spectra for events that start as a primary excitation of the $\Delta$ resonance. Even though their cross section is smaller than that for single-pion production its shape is very similar. This reflects the fact that the spectra are more determined by the fsi than the primary excitation process.

Different from the case at the lower energies is the presence of a long, rather significant tail in the spectrum out to higher pion kinetic energies. This is due to the initial production of higher energy pions (mainly through DIS) which cannot cascade down all the way into the peak region. Furthermore, at the higher energies, above about 0.5 GeV, the spectra before and after fsi are very similar. Overall, as a net effect, pion absorption is less pronounced at the MINER$\nu$A energies than it was at MINIBooNE/T2K. This is mainly a consequence of the minimum in the $\pi-N$ cross section at around 0.7 GeV kinetic energy.

\subsubsection{Kaon Production}
\begin{figure}
\includegraphics[angle=-90,width=0.5\textwidth]{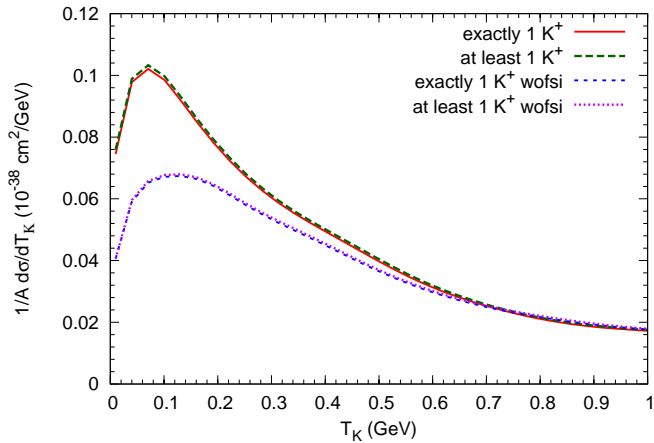}
\caption{(color online) Flux-averaged kinetic energy distribution of $K^+$s. The uppermost solid (red) curve gives the cross section per nucleons for events with at least 1 $K^+$ and any number of other hadrons, the dashed (green) curve that for events with exactly 1 $K^+$ and no other kaons. The other 2 curves give the corresponding results without any final state interactions.} \label{fig:K+}
\end{figure}

\begin{figure}
\includegraphics[angle=-90,width=0.5\textwidth]{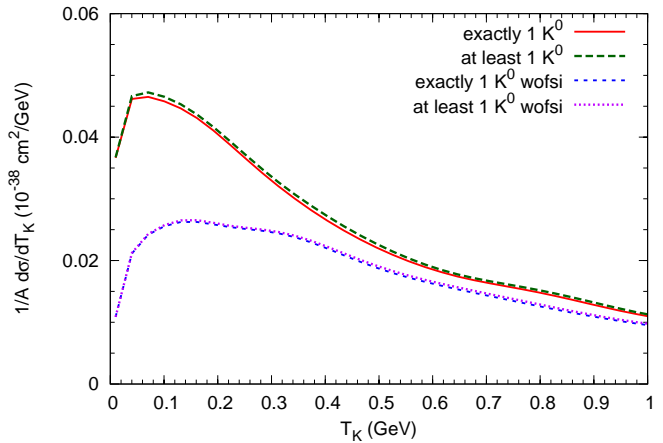}
\caption{(color online) Flux-averaged kinetic energy distribution of $K^0$s. The uppermost solid (red) curve gives the cross section per nucleons for events with at least 1 $K^0$ and any number of other kaons, the dashed (green) curve that for events with exactly 1 $K^0$ and no other kaons. The other 2 curves give the corresponding results without any final state interactions.} \label{fig:K0}
\end{figure}

MINER$\nu$A has also plans for investigations of strangeness production \cite{Solomey:2005rs}. To provide some guidance for these experiments we show in Figs.\ \ref{fig:K+} and \ref{fig:K0} the kinetic energy spectra of both positively charged $K^+$ and neutral $K^0$ kaons because both have the same strangeness content. Both spectra peak at about 0.1 GeV and then fairly slowly fall off towards higher energies. The cross sections for exclusive single and semi-inclusive $K$ production hardly differ, in contrast to the situation for pion production.

Interesting is the comparison between the spectra after fsi (upper 2 curves) and those before fsi (lower 2 curves). In striking contrast to the situation for pion production the fsi now \emph{increase} the cross section for all kaon energies by factors of about 1.5 - 1.8, for $K^+$ and $K^0$, respectively. This increase is due the fact that these kaons can be created in secondary collisions, such as $\pi + N \to K + \Lambda$. At the same time they have a long mean free path in nuclear matter because of strangeness conservation.

\section{Discussion and Conclusions}
Over the last few years a lot of emphasis has been put on the correct description of QE scattering, including 2p-2h processes. Significantly less theoretical effort has been spent on resonance excitations, pion production and DIS. This is surprising since experimentally these processes are strong, cannot be distinguished from each other or from QE scattering and are intimately entangled with the latter. As a consequence, the extraction of QE processes (true QE and 2p-2h) requires the use of a generator and thus introduces uncertainties into the measured cross sections. At MINER$\nu$A, with its higher energy, this is even more so than at the lower-energy experiments MiniBooNE and T2K. At higher energies more processes are active and their strength depends on neutrino energy. For example, the results shown here make it extremely difficult to isolate the 2p-2h contributions at the higher energies of the MINER$\nu$A experiment. This process disappears in a background from higher resonance excitations, DIS and pion background processes (see Fig.\ \ref{fig:Q2nopi}), all with similar final states.  The 2p-2h contribution is generally quite small in the flux-averaged final 0 pion event sample, compared to true QE and pion production channels. Thus the sensitivity to details of the treatment of 2p-2h contributions is smaller than the uncertainties introduced by the $Q^2$ reconstruction and our insufficient knowledge of pion production cross sections.

A general problem arises if experimental flux distributions are being cut in the data analysis. For example, a low energy cut-off in the MINER$\nu$A flux was introduced in the experimental work of Refs.\ \cite{Fiorentini:2013ezn,Fields:2013zhk} mainly to take care of the fact that MINER$\nu$A can see no muons below about 1.5 GeV, since it is using the forward MINOS spectrometer. The comparison of the event rates in Fig.\  \ref{fig:ev_no_pion} shows, however, that this cut-off introduces a problem into the analysis. The acceptance limitation is one in terms of true neutrino energy and, therefore, in the calculations presented here the incoming neutrino flux has been cut at the \emph{true} energy of 1.5 GeV. In the experimental analysis, however, that energy is not directly accessible, but must be reconstructed. The only cut an experimental analysis can introduce is one on the \emph{reconstructed} neutrino energy. Even if in Fig.\ \ref{fig:ev_no_pion} such a cut at 1.5 GeV reconstructed energy is performed there remains quite a significant difference between the two event rates: cutting the flux below the reconstructed energy of 1.5 GeV removes sizeable strength at higher true energies (see Fig.\ \ref{fig:ev_no_pion}) and distorts the event rate.

The cut-off thus introduces an error due to an imperfect energy reconstruction into the data: the data become model-dependent, the model being here Eq.\ (\ref{E_Q2}) together with the restriction to 0-pion events. This is even more so if in addition other event classes are removed from the full event sample by means of a generator; in this case the 'data' are both model- and generator-dependent\footnote{We distinguish here between a model and a generator. The 'model' consists of a clean-cut prescription that relies on experimentally observable quantities only. The 'generator' contains theoretical input and prescriptions for processes that cannot directly be accessed experimentally.}. Any theoretical analysis then runs immediately into inconsistencies since the 'data' contain effects of one (possibly less than perfect) model and generator while a 'perfect' theory may employ another description of the reaction mechanisms.
In such a case it is difficult to see if differences between theoretically calculated and experimentally derived observables are due to difficulties of the underlying theory or just to weaknesses in the generators used in the experimental analysis.

Therefore, the influence of models and generators on the data should be minimized by as much as possible. While the reconstruction of the true energy always requires a generator, the extra effects coming from modeling the flux cut-offs should be avoided. In the light of these considerations the comparison of the extracted QE-like data with different models, presented in \cite{Fiorentini:2013ezn}, has to be reconsidered: about 75\% of the data points used for that discussion lie in the region ($Q^2 < 0.6$ GeV$^2$) where the reconstruction effects are largest (see Fig.\ \ref{fig:Q2}). Similar remarks as for QE also apply to data on pion production that are presently being analyzed at MINERV$\nu$A \cite{Dytmanprivcom:2013}. Also here low- and high-energy cuts have an effect on the cross sections.
Cuts on the flux should thus be avoided; a superior strategy would be to provide reasonably accurate acceptance filters (in the case of MINER$\nu$A) for the outgoing muon to be implemented in event generators.

The problems just discussed arise because nuclear fsi make it impossible to uniquely identify QE scattering. These fsi can also shield the elementary interaction vertices. For MINER$\nu$A this is important since this experiment was motivated by the need for better neutrino-nucleon cross sections. We have shown that knock-out nucleons are strongly affected by fsi also in the MINER$\nu$A experiment. In \cite{Fiorentini:2013ezn} a threshold energy of 225 MeV was used to search for evidence for multinucleon processes. Our results indicate that at this energy and below there is a strong effect of 'normal' fsi. The 0 pion constraint redistributes energy from baryons above to those below just this energy.

It is well known that pions are strongly absorbed in nuclei. At the MINER$\nu$A energies this is more an effect on the shape of kinetic energy distributions and not so much on the total yield. At energies around about 80 MeV due to fsi the pion yield even gets higher than the cross section without any fsi. Furthermore, the higher-energy pions above a few hundred MeV kinetic energy are only weakly affected by fsi. Any new data on pion production from MINER$\nu$A could thus be very helpful in answering the open questions on the correct elementary pion production cross section and on the pion fsi. We note, however, that any cuts on the invariant mass $W$ for pion production introduce another uncertainty into the data since $W$ again has to reconstructed by using the reconstructed energy and $Q^2$. Pion data so obtained will suffer from the same problems as the event and $Q^2$ distributions discussed earlier.

The effects of fsi are different for kaon production. At first sight, both $K^+$ and $K^0$ seem to be promising for an investigation of elementary kaon production cross sections because these mesons, due to their special strangeness content, are expected to suffer only little absorption. However, our results show that there is a strong contribution to the kaon yield from secondary reactions in the nuclear medium in which mainly pions contribute to the kaon yield. These fsi processes \emph{increase} the strangeness production cross section for all kaon kinetic energies. This strong effect of final state interactions makes it thus very difficult to actually determine the elementary neutrino-induced kaon production cross sections when nuclear targets are used.

Both of the results discussed in this paper, the problems connected with energy and $Q^2$ reconstruction and the influence of the nuclear medium on the extraction of elementary meson production cross sections, highlight the need for a state-of-the-art generator. The description of inclusive cross sections is not sufficient; the all-important final state interactions must be described with the same level of sophistication up to the final, asymptotic state.  The 'precision era' of neutrino physics also requires 'precision era' generators!

\begin{acknowledgments}
We thank S. Dytman for providing us with the most recent version of the MINER$\nu$A flux. We also thank L. Fields and K. McFarland for many helpful comments on the actual data analysis in MINER$\nu$A and are grateful to K. McFarland for comments on the first version of this paper. Finally, we gratefully acknowledge the help and support of the whole GiBUU team in developing both the physics and the code used here.

This work has been supported by BMBF and LOEWE.
\end{acknowledgments}

\bibliographystyle{apsrev4-1}
\bibliography{nuclear}

\end{document}